\documentstyle{article}
\begin{document}
\author{V.~Ya.~Demikhovskii and A.~A.~Perov}
\title{Hall Conductance of a Two-Dimensional Electron Gas with Spin-Orbit
Coupling at the Presence of Lateral Periodic Potential}
\maketitle

\vspace{0.5cm}

Nizhny Novgorod State University, Gagarin ave., 23, Nizhny Novgorod 603950, Russia

\vspace{0.5cm}
\begin{abstract}
We evaluate the distribution of Hall conductances in magnetic subbands
of two-dimensional electron gas with Rashba spin-orbit (SO) coupling
placed in a periodic potential and perpendicular magnetic field. In this
semiconductor structure the spin-orbit coupling mixes the states of different
magnetic subbands and changes the distribution of their Hall conductances
in comparison with the case of spinless particles. The calculations were
made for semiconductor structures with a weak ($AlGaAs/GaAs$) and
relatively strong ($GaAs/InGaAs$) SO and Zeeman interactions. The Hall
conductances of fully occupied magnetic subbands depend on the system
parameters and can be changed when neighboring subbands touch each other.
It was shown that in the real semiconductor structures with relatively
strong SO coupling the distribution of Hall conductance differs from
the quantization law predicted by Thouless et al\cite{Thoul} for systems without
spin-orbit coupling. In the case of weak SO interaction and relatively large
lattice period the Hall conductance of magnetic subbands are the same as
for spinless particles, but as the lattice period decreases and two neighboring
subbands touch each other the distribution of Hall conductances is changed
drastically.
\end{abstract}

\vspace{0.5cm}

\vspace{0.5cm}
\section{Introduction}
During last decade the intricate energy spectrum of two-dimensional electrons
subject to both periodic potential and a perpendicular magnetic field, has
been studied theoretically and experimentally\cite{{Sch},{KvK},{Geis}}. The first
experimental observation of the splitting of magnetic levels into subbands in
an $n$-type heterojunction at the presence of a surface superlattice was
observed for the first time in\cite{Sch}. The effects of magnetic subbands
formation have also been studied in\cite{{KvK},{Geis}}, where the
magnetotransport in the structures with a surface superlattice was
investigated. However, the analysis of the experimental results was based on
theoretical models which do not take into account the effects of an electron
spin-orbit coupling.

In a profound paper by Thouless, Kohmoto, Nightingale and den Nijs\cite{Thoul}
an explicit formula for the Hall conductance of 2D noninteracting electrons in
a periodic potential was derived. As it was found in\cite{Thoul}, such a
system has integral values of the Hall conductance in units of $e^2/h$ if the
Fermi energy lies in a gap between magnetic subbands. This result has a
topological nature and, as it was mentioned in many works, does not depend on
the detailed structure of the potential\cite{Us}.

The theoretical studies of the magnetotransport in 2D electron gas in the
presence of SO coupling demonstrate that in high magnetic field where
the integer quantum Hall effect is observed, a sufficiently strong Rashba
constant creates new plateau in the Hall resistivity and splits the SdH
peaks of $\rho_{xx}$\cite{Vas}. Later\cite{Shen} the spin Hall conductance of
2D electron gas with Rashba SO coupling was calculated. In particular,
the resonance increase of spin Hall conductance was predicted when
two energy levels cross each other at the Fermi level.

In our previous work\cite{DP} the Harper-Hofstadter problem for two-dimensional
SO coupled system subject to a periodic potential and perpendicular
magnetic field was investigated. We have analyzed analytically and
numerically the magnetic Bloch states of 2D electrons subject to both periodic
potential of a lateral superlattice and perpendicular magnetic field under the
conditions when the SO coupling and Zeeman term are taken into consideration.
Also, the spin density distribution in the elementary cell is obtained and the
average spin polarization in a state with given quasimomentum is calculated.

In this paper we investigate the quantum Hall effect in 2D electron gas with
SO Rashba coupling arising from inversion asymmetry subject to both periodic
potential of a superlattice and a uniform perpendicular magnetic field. The
calculations were carried out for two semiconductor systems with weak and quite
strong SO interaction. The Zeeman effect for electrons was also taken into
account. As was established, the Hall conductance quantization rule depends on
the system parameters at fixed magnetic flux per unit cell of the superlattice
and changes at the moment when the neighboring magnetic subbands touch each
other. As we will see below, in the system with a weak SO coupling when the SO
and Zeeman splittings are smaller than Landau level splitting due to the
action of periodic potential the neighboring magnetic subbands are overlapped
but not degenerated for any quasimomentum ${\bf k}$. In that case the Hall
conductance quantization rule is the same as for spinless
particles\cite{Thoul}. We found that with the decreasing superlattice period
the effect of SO coupling rises and the distribution of Hall conductances
in magnetic subbands is changed at the moment when neighboring subbands touch
each other. In semiconductor structures where SO coupling as well as Zeeman
splitting are larger than the splitting by periodic potential, we found a new
distribution law for Hall currents of fully occupied magnetic subbands.

All calculations performed in this paper have been made without taking into
account the potential disorder. We assume that, just as in the absence of the
spin-orbit coupling, the presence of disorder gives rise to localization of
quantum states. So, the extended states in neighboring Landau levels are
separated in energy by localized states.

The paper is organized as follows. Section II is devoted to the calculations
of energy spectrum and spinor Bloch wave functions of an electron in two realistic
semiconductor superlattice structures placed in uniform magnetic field. The
position of Landau subbands versus the magnetic field and dispersion laws
when the number of magnetic flux quanta per unit cell of the superlattice
is equal to three flux quanta are evaluated for these semiconductor systems
with different strengths of Rashba and Zeeman interactions. In Section III
we analyse the structure of Berry curvatures which are defined in the magnetic
Brillouin zone (MBZ) and discuss the numerical results for Hall conductance
distributions for both cases of weak and strong SO coupling.

\section{Quantum states}
We consider 2D electron gas with spin-orbit Rashba coupling in a potential
$V(x,y)$ which is periodic in plane with the period $a$, and in a uniform
magnetic field ${\bf H}$ perpendicular to the plane of the  electrons. The
correspondent one-electron Hamiltonian has the following form:
$$
\hat H=\hat H_0 + V(x,y),\eqno(1)
$$
where $V(x,y)=V(x+na,y+ma)$,
$$
\hat H_0=({\bf\hat p}-e{\bf A}/c)^2/2m^{\ast}+\hat H_R-g\mu_B H\hat\sigma_z,
$$
$$
\hat H_R=\frac{\alpha}{\hbar}\bigg(\hat\sigma_x (\hat p_y-eA_y/c)-
\hat\sigma_y\hat p_x\bigg)
$$
is the Rashba Hamiltonian of an electron in uniform magnetic field. Here
$\hat p_{x,y}$ are the momentum operator components, $m^{\ast}$ is the
electron effective mass, $\hat\sigma_i\, (i=x,y)$ are the Pauli matrices,
$\alpha$ is the parameter of the SO coupling, $g$ is  the  Zeeman factor, and
$\mu_B$ is the Bohr magneton. We use the Landau gauge in which the vector
potential has the form ${\bf A}=(0,Hx,0)$ and consider the potential
$V(x,y)=V_0(\cos(2\pi x/a)+\cos(2\pi y/a))$. The quantum states structure of
the system under consideration depends crucially on the parameter
$\Phi/\Phi_0=p/q=|e|Ha^2/2\pi hc$ ($p$ and $q$ are prime integers) which
is the number of flux quanta per unit cell, and $\Phi_0$ is the flux quanta.

One can express the eigenfunction of Hamiltonian (1) as a set of Landau wave
functions in the presence of the SO coupling. If we take $p/q$ to be the
rational number, the two-component magnetic Bloch function, analogous to the
one-component magnetic Bloch function of the $\mu$th magnetic subband, can be
written in the form
$$
\displaylines{\Psi_{{\bf k},\mu}(x,y)=\pmatrix{\Psi_{1{\bf k},\mu}(x,y)\cr
\Psi_{2{\bf k},\mu}(x,y)}=\sum\limits_{n=1}^{p}
\sum\limits_{\ell=-\infty}^{+\infty}{\rm e}^{ik_x(\ell qa+nqa/p)}
{\rm e}^{2\pi iy(\ell p+n)/a}\times\cr
\hfill\times\Bigg[A_{0n}^{\mu}({\bf k})\psi_{0n\ell {\bf k}}^{+}(x,y)+
\sum\limits_{S=1}^{\infty}\{ A_{Sn}^{\mu}({\bf k})\psi_{Sn\ell
{\bf k}}^{+}(x,y)+B_{Sn}^{\mu}({\bf k})\psi_{S,n\ell {\bf k}}^{-}(x,y)\}
\Bigg],\hfill\llap{(2)}\cr}
$$
where the spinors $\psi_{0n\ell {\bf k}}^{+}=\exp(ik_yy)\pmatrix{0\cr
\phi_0[\xi_{\ell n}]}$, $\psi_{Sn\ell {\bf k}}^{+}=\frac{\exp(ik_yy)}
{\sqrt{1+D_S^2}}\pmatrix{D_S\phi_{S-1}[\xi_{\ell n}]\cr
\phi_S[\xi_{\ell n}]}$ and $\psi_{Sn\ell {\bf k}}^{-}=\frac{\exp(ik_yy)}
{\sqrt{1+D_S^2}}\pmatrix{\phi_{S-1}[\xi_{\ell n}]\cr
-D_S\phi_S[\xi_{\ell n}]}$ correspond to "$+$" and "$-$" branches of the
spectrum of the Hamiltonian $\hat H_0$\cite{Vas}. Here,
$D_S=(\sqrt{2S}\alpha/l_H)/(E_0^++\sqrt{(E_0^+)^2+2S\alpha^2/l_H^2})$,
$\phi_S[\xi]$ are the simple harmonic oscillator functions,
$l_H=c\hbar/|e|H$ is the magnetic length, $E_0^+=\hbar\omega_c/2+g\mu_BH$
and $\omega_c=|e|H/m^{\ast}c$ is the cyclotron frequency. Here quantum numbers
$S=1,2,3,\ldots$ characterize the pair of "$+$" and "$-$" states
$E_S^{\pm}=S\hbar\omega_c\pm\sqrt{(E_0^+)^2+2S\alpha^2/l_H^2}$ of the
unperturbed Hamiltonian $\hat H_0$, $\xi_{\ell n}=(x-x_0-\ell qa-nqa/p)/l_H$,
$x_0=c\hbar k_y/|e|H$.

Note that the spinor wave function (2) is the eigenfunction of both the
Hamiltonian (1) and the operator of magnetic translation and therefore it has
to obey the following Bloch-Peierls conditions
$$
\Psi_{{\bf k},\mu}(x+qa,y+a)=\Psi_{{\bf k},\mu}(x,y)\exp(ik_xqa)\exp(ik_ya)
\exp(2\pi iy/a),
$$
where ${\bf k}$ is the quasimomentum defined in the MBZ
$$
-\pi/qa\le k_x\le\pi/qa,\quad -\pi/a\le k_y\le\pi/a.
$$
So, here the magnetic Brillouin zone is the same as for the charged spinless
particle. In the limit of high magnetic fields the functions
$\psi_{Sn\ell {\bf k}}^+$ and $\psi_{Sn\ell {\bf k}}^-$ are proportional to
eigenspinors of Pauli operator $\hat\sigma_z$: $\pmatrix{1\cr 0}$ and
$\pmatrix{0\cr 1}$.

Substituting Eq.(2) in the Schr\"odinger equation $\hat H\Psi=E\Psi$ we come
to the infinite system of linear equations for coefficients
$A_{Sn}^{\mu}({\bf k})$ and  $B_{Sn}^{\mu}({\bf k})$. In the case when periodic
potential amplitude and spin-orbit coupling energy have the same order and the
inequality $\Delta E_{SO}\simeq V_0\le \hbar\omega_c$ takes place the system of
linear equations can be reduced to the system of infinite number of uncoupled
groups of $2p$ equations. Each group describes the magnetic Bloch states formed
from the states with energies $E_S^+$ and $E_{S+1}^-$ of the single Landau
level split by SO interaction. The structure of the Hamiltonian matrix as well
as the matrix elements have been discussed in details in our previous
paper\cite{DP}. It was demonstrated that SO interaction mixes the states of
pure Landau levels and results in a doubling of the number of the magnetic
subbands formed under the fixed value $p/q$ of magnetic flux quanta per unit
cell.

The experimental values of Rashba coupling constant $\alpha$ for different
materials range from about $2\cdot 10^{-12}\, {\rm eV}\cdot {\rm m}$ to
$4\cdot 10^{-11}\, {\rm eV}\cdot {\rm m}$.
The calculations of quantum states were carried out for two semiconductor
structures with different strength of SO coupling. One of them is 2D
electron gas in $AlGaAs/GaAs$ heterojunction with superlattice periodic
potential. This structure is characterized by small SO Rashba constant $\alpha$
and relatively small $g$-factor. The parameters of another system correspond
to $GaAs/In_x Ga_{1-x} As$ heterostructure where SO coupling parameter and
$g$-factor have large values.

We have calculated the position of magnetic subbands at different $p/q$ as
well as the dispersion laws $E_{\mu}({\bf k})$ in magnetic Brillouin zone.
In Figs.1a, 1b the distribution of magnetic subbands versus the
number of magnetic flux quanta per unit cell $p/q$ is presented. The spectrum
shown in Fig.1a corresponds to the parameters of $AlGaAs/GaAs$ structure,
and magnetic subbands in $GaAs/In_x Ga_{1-x} As$ system are presented in
Fig.1b. The magnetic subbands split off the four lowest Landau levels are
shown in both cases. The arrows mark the position of energy bands for the
value $p/q=3/1$ at which the calculations of Hall conductance were carried
out (see below).


The dispersion laws $E_{\mu}({\bf k})$ in MBZ are presented in Fig.2a,
2b for $p/q=3/1$ for parameters $AlGaAs/GaAs$ and $GaAs/In_{0.23}Ga_{0.77}As$
structures, respectively. Here and below we consider the magnetic subbands
attached to the lowest pair of unperturbed levels $E_0^+$ and $E_1^-$.
The system parameters here are the same as in Fig1. The positions of these
six ($2p$) subbands are pointed out by arrows at Fig.1a, 1b. As one can see
from Fig.1a and Fig.2a the Landau level is split off into magnetic subbands
due to the periodic potential of the lattice. The splitting caused by the SO
coupling and Zeeman effect is quite small because of the small Rashba constant
and Lande $g$-factor in the $GaAs$. Thus, the spin degeneracy of an electron
vanishes and three pairs of overlapped magnetic minibands are formed
(see Fig.2a). So, in Fig.1a these six subbands are visible as three bands. It
should be noted, that the ratio of the SO and Zeeman splittings to the level
broadening produced by a periodic potential essentially depends on the lattice
period $a$. We calculated also the energy spectrum consisting of six magnetic
subbands ($p/q=3/1$) for the system $AlGaAs/GaAs$ with superlattice periods
$a<80\, {\rm nm}$. It was observed that with the decreasing superlattice period
all magnetic subbands become not overlapped. Note also, that at the period
$a\simeq 56\, {\rm nm}$ the second and the third subbands touch each other and
after that all subbands are separated by gaps again. As we will see below
such a reconstruction of energy spectrum leads to the changing of Hall
conductance quantization laws in fully occupied Landau subbands.

In contrast, in $GaAs/In_{0.23}Ga_{0.77}As$ structure all of six magnetic
subbands are not overlapped due to the large SO interaction and Zeeman
splitting of the unperturbed discrete levels $E_0^+$ and $E_1^-$. As we will
see below, the existence of small energy gaps in the spectrum leads to the
specific shape of Berry curvature which defines the Hall conductance
quantization law of fully occupied magnetic subbands.


\section {Hall conductance}
The quantum Hall effect in 2D periodic potential has a topological nature.
We have investigated the influence of the spin-orbit coupling on topological
invariants (first Chern numbers) of magnetic subbands, which determine their
Hall conductance.

When a small electric field is applied, a resulting current in the
perpendicular direction to applied electric field may be given by Kubo formula.
So, the resulting Hall conductance of the occupied magnetic subbands can
be written as
$$
\sigma_{xy}=-ie^2\hbar\sum\limits_{E_\nu({\bf k})<E_F<E_\mu({\bf k})}
\frac{(v_y)_{\nu{\bf k},\mu{\bf k}}(v_x)_{\mu{\bf k},\nu{\bf k}}-
(v_x)_{\nu{\bf k},\mu{\bf k}}(v_y)_{\mu{\bf k},\nu{\bf k}}}
{(E_\nu({\bf k})-E_\mu({\bf k}))^2},\eqno(3)
$$
where $E_F$ is a Fermi energy and the summation implies the sum over all the
states below and above the Fermi level. The indices $\nu$ and $\mu$ label
magnetic subbands. At the presence of SO coupling the components of velocity
operator in $x$ and $y$ directions $v_x$ and $v_y$ are defined by the
expressions:
$$
v_x=\frac{\partial H}{\partial p_x}=
\pmatrix{-i\hbar\nabla_x/m^\ast & i\alpha/\hbar \cr
-i\alpha/\hbar & -i\hbar\nabla_x/m^\ast}\eqno(4.a)
$$
$$
v_y=\frac{\partial H}{\partial p_y}=
\pmatrix{-i\hbar\nabla_y/m^\ast-\omega_cx & \alpha/\hbar \cr
\alpha/\hbar & -i\hbar\nabla_y/m^\ast-\omega_cx}.\eqno(4.b)
$$

Using Eq.(3) after simple transformations one can write the Hall
conductance of the $\mu$th fully occupied magnetic subband in the following
form\cite{{Us},{CN}}:
$$
\sigma_{xy}^{\mu}=\frac{e^2}{2\pi h}\int\Omega_{\mu}({\bf k})\, d^2{\bf k}=
\frac{e^2}{\hbar}\int \bigg[-l_H^2+\Omega_{\mu}^{(1)}({\bf k})\bigg]\,
d^2{\bf k}.\eqno(5)
$$
In the representation of the Hamiltonian $\hat H_0$ the following expression
for $\Omega_{\mu}^{(1)}({\bf k})$ takes place:
$$
\Omega_{\mu}^{(1)}({\bf k})=i\sum\limits_{\nu\ne\mu}\Bigg[
\frac{\bigg(\sum\limits_{mm^\prime}d_m^{\mu\ast}({\bf k})
d_{m^\prime}^{\nu}({\bf k})\frac{\partial H_{mm^\prime}}{\partial k_x}\bigg)
\cdot
\bigg(\sum\limits_{ll^\prime}d_{l^\prime}^{\nu\ast}({\bf k})
d_l^{\mu}({\bf k})\frac{\partial H_{l^\prime l}}{\partial k_y}\bigg)}
{(E_\mu({\bf k})-E_\nu({\bf k}))^2}-{\rm c.c.}\Bigg].
\eqno(6)
$$
Here, the matrix elements of the Hamiltonian matrix $H_{mm^\prime}$ are defined
in our previous paper\cite{DP}. The coefficients $d_m({\bf k})$ are the
eigenvectors of the Hamiltonian matrix
$H_{mm^\prime}$ ($m,m^\prime=1,\ldots,2p$), related to the coefficients
$A_{Sn}({\bf k})$ and $B_{Sn}({\bf k})$ in Eq.(2).

In Eq.(5) the integration is over the occcupied states in the MBZ, and the
function $\Omega_{\mu}({\bf k})$ is the $z$-component of the Berry curvature
whose integral over an area bounded by a path $C$ in ${\bf k}$-space is the
Berry phase $\gamma_{\mu}(C)$\cite{Ber}.
The $z$-component of the Berry curvature is defined as usial by the
expression\cite{Thoul}
$$
\Omega_{\mu}({\bf k})=i\Bigg(\bigg<\frac{\partial u_{{\bf k},\mu}}{\partial k_x}
\frac{\partial u_{{\bf k},\mu}^{\ast}}{\partial k_y}\bigg>-c.c.\Bigg).\eqno(7)
$$
In our case $u_{{\bf k},\mu}(x,y)$ is the periodic part of spinor
magnetic Bloch function (2):
$u_{{\bf k},\mu}(x,y)=\Psi_{{\bf k},\mu}(x,y)\cdot\exp (-i{\bf k}{\bf r})$.
The first term in square brackets in Eq.(5) defines the "ideal" Hall
conductance of a single magnetic subband\cite{Us} equals $e^2/ph$.
At the same time due to Laughlin arguments\cite{La} each magnetic subband with
its extended states completely filled contribute whole multiple of $e^2/h$ to
the Hall conductance. So, the integral of $\Omega_{\mu}({\bf k})$ over one
magnetic Brillouin zone divided by $2\pi$ is always integer which is the
topological Chern number.

The semiclassical dynamics of magnetic Bloch electrons was investigated by
Chang and Niu\cite{CN}. Using the magnetic Bloch states they derived the
semiclassical equations of electron motion and showed that the Berry curvature
of magnetic band plays a crucial role in the electron subband dynamics. Namely,
it gives electrons an extra velocity in the direction of
${\bf E}\times {\bf H}$ and defines the Hall conductivity quantization law.
We have made a considerable use of these results\cite{CN}.

The shapes of the functions $\Omega_{\mu}({\bf k})$ for
different magnetic subbands are displayed in Figs.3a, 3b for $AlGaAs/GaAs$ and
$GaAs/InGaAs$ structures, respectively. As one can see from the comparison
of Figs.2a, 2b and Figs.3a,3b the maxima of $\Omega_{\mu}({\bf k})$
are located in the ${\bf k}$-space along the lines where the gaps between
neighbor magnetic subbands are narrow.



The distribution of Hall conductance in magnetic subbands for $AlGaAs/GaAs$
and $GaAs/InGaAs$ superlattice structures is shown in Fig.4a and Fig4b,
respectively. Here the magnetic flux through the superlattice elementary cell
is equal to three flux quanta. In systems with weak SO coupling (see Fig.1a
and Fig.2a) when the SO and Zeeman terms are smaller than the Landau level
broadening produced by the periodic potential, a quant of the Hall current is
carried by the middle subband from each group of three subbands (Fig.4a). So,
the results of our calculations are in good agreement with
Thouless et al.\cite{Thoul} results that in a weak periodic potential and at
the absence of spin-orbit and Zeeman interactions only the central subband of
each Landau level carries the Hall current when $q=1,3,5,\ldots$. With the
decreasing superlattice period the distribution of Hall conductances of
magnetic subbands is changed drastically. As was mentioned above at the
critial value of the period $a=56\, {\rm nm}$ the second and the third subbands
contact each other and the dispersion laws $E_{\mu}({\bf k})$ ($\mu=2,3$) are
modified. At the same time the correspondent Berry curvatures
$\Omega_{\mu}({\bf k})$ and, therefore, the topological invariants (first
Chern numbers) of these subbands defining their Hall conductance are changed
dramatically. We found that in $AlGaAs/GaAs$ superlattice structure at
$p/q=3/1$, $a=50\, {\rm nm}$, $V_0=1\, {\rm meV}$, $g=-0.44$ and
$\alpha=2.5\cdot 10^{-12}\, {\rm eV\cdot m}$ the following sequence of Hall
conductances in units of $(-e^2/h)$ is realized: 0,2,-1,1,0,0. As a result,
in the system which is characterized by a weak Rashba and Zeeman coupling
constants at the certain lattice parameters the distribution of Hall
conductances can be totally different from another one distribution predicted
for spinless particles\cite{Thoul}.

If the SO and Zeeman terms are comparable to
the Landau level broadening caused by the superlattice periodic potential, the
distribution of Hall conductance across six subbands measured in units of
$-e^2/h$ follows the same sequence as before: 0, 2, -1, 1, 0, 0 (see Fig.4b).
Correspondingly, the Hall current here is carried by the second, the third,
and the fourth magnetic subbands. Moreover, we observed that each time when
two bands touch each other while the SO or Zeeman coupling parameters are
varied, the conductances of these subbands may not be conserved when the bands
split again. However, the sum of the Hall conductances remains unchanged.

\section{Conclusions}
To conclude, we have investigated the Hall conductance distributions in 2D
superlattice system with Rashba spin-orbit coupling at the presence of
perpendicular magnetic field. The electron spinor wave function is the
eigenfunction of both the Hamiltonian and operator of magnetic translation and
obeys the Bloch-Peierls conditions. The quantum states for the structures with
a weak and relatively strong SO and Zeeman interactions are evaluated. It was
demonstrated that in the case of weak SO interaction when the Landau level
broadening due to a periodic potential is large than Rashba and Zeeman terms,
the overlapped magnetic  subbands are grouped into pairs. In the opposite case
all magnetic subbands are resolved and separated by energy gaps.

The Hall conductances of magnetic subbands were obtained by integration of
the Berry curvatures over magnetic Brillouin zone. It was found that in real
semiconductor superlattice systems subject in uniform magnetic field at the
presence of spin-orbit coupling the distribution of Hall conductances in
magnetic subbands can differ from another one obtained by Thouless et
al\cite{Thoul} in the model of spinless particles. Even in the case of
relatively weak SO coupling and Zeeman term the distribution of Hall
conductances is changed with the decreasing lattice period.

\section*{Acknowlegment}
This work was supported by the program of the Russian Ministry of Education
and Science "Development of scientific potential of high education" (project
2.1.1.2363), grant of Russian Foundation of Basic Research (no. 06-02-17189)
and grant of the President of Russian Federation (MK-5165.2006.2).

\end{document}